\documentclass{PoS}

\usepackage{bm}
\usepackage{graphicx}% Include figure files
\usepackage{amsmath, amssymb}% for complicated equations

\newcommand{\Psfig}[2]{\includegraphics[width=#1]{#2}}

\def\gev{\text{ GeV}}
\def\fm{\text{ fm}}

\title{Size measurement of dynamically generated hadronic resonances with finite volume effect}

\ShortTitle{Size measurement of dynamically generated hadronic resonances}

\author{\speaker{Takayasu Sekihara}\\
         Institute of Particle and Nuclear Studies, High
         Energy Accelerator Research Organization (KEK), 1-1, Oho, Ibaraki
         305-0801, Japan\\
         E-mail: \email{sekihara@post.kek.jp}}

\author{Tetsuo Hyodo\\
        Yukawa Institute for Theoretical Physics, 
        Kyoto University, Kyoto 606-8502, Japan\\
        E-mail: \email{hyodo@yukawa.kyoto-u.ac.jp}}

      \abstract{The structures of the hyperon resonance $\Lambda
        (1405)$ and the scalar mesons $\sigma$, $f_{0}(980)$, and
        $a_{0}(980)$ are investigated based on the coupled-channels
        chiral dynamics with finite volume effect.  The finite volume
        effect is utilized to extract the coupling constant,
        compositeness, and mean squared distance between two
        constituents of a Feshbach resonance state as well as a stable
        bound state. In this framework, the real-valued size of the
        resonance can be defined from the downward shift of the
        resonance pole according to the decreasing finite box size $L$
        on a given closed channel.
        As a result, we observe that, when putting the $\bar{K}N$ and
        $K\bar{K}$ channels into a finite box while other channels
        being unchanged, the poles of the higher $\Lambda (1405)$ and
        $f_{0}(980)$ move to lower energies while other poles do not
        show downward mass shift, which implies large $\bar{K}N$ and
        $K\bar{K}$ components inside higher $\Lambda (1405)$ and
        $f_{0}(980)$, respectively.  Extracting structures of $\Lambda
        (1405)$ and $f_{0}(980)$ in our method, we find that the
        compositeness of $\bar{K}N$ ($K\bar{K}$) inside $\Lambda
        (1405)$ [$f_{0}(980)$] is $0.82$ -- $1.03$ ($0.73$ -- $0.97$)
        and the mean distance between two constituents is evaluated as
        $1.7$--$1.9 \fm$ ($2.6$--$3.0 \fm$).  }

\FullConference{XV International Conference on Hadron Spectroscopy \\
                 4-8/11/2013\\
                 Nara, Japan}

\begin{document}

\section{Introduction}

Establishing exotic hadrons, which have different quark configurations
from the ordinary $q\bar{q}$ (mesons) or $qqq$ (baryons) state, is one
of the most important issues in the physics of strong interaction.
This is because there is no clear experimental evidence on the
existence of the exotic hadrons while quantum chromodynamics, the
fundamental theory of strong interaction, does not prohibit their
existence.  A classical example of the exotic hadron candidates is
$\Lambda (1405)$, which has been considered as an $s$-wave $\bar{K}N$
quasi-bound state rather than an $uds$ three-quark
state~\cite{Dalitz:1960}.  The lightest scalar mesons in a nonet
state, $\sigma = f_{0}(500)$, $\kappa = K_{0}^{\ast}(800)$,
$f_{0}(980)$, and $a_{0}(980)$ are also candidates of the exotic
hadrons, and they may be multi-quark hadrons~\cite{Jaffe:1976ig} or
$K\bar{K}$ quasi-bound states for $f_{0}(980)$ and
$a_{0}(980)$~\cite{Weinstein:1982gc}.  Recently both $\Lambda (1405)$
and the lightest scalar mesons have been well described
phenomenologically in the so-called chiral unitary approach in the
meson-baryon~\cite{Kaiser:1995eg} and meson-meson~\cite{Dobado:1992ha}
scatterings, respectively.  To pin down the structure of these
hadrons, we here focus on their spatial size.

If a resonance state is a quasi-bound state with a small binding
energy, one can expect that the resonance state has a larger spatial
size than typical size of the constituents. Motivated by this
expectation, the spatial structures of $\Lambda (1405)$ and $\sigma$
are theoretically measured in the meson-baryon~\cite{Sekihara:2008qk}
and meson-meson~\cite{Albaladejo:2012te} scatterings, respectively,
and it is found that the baryonic and strangeness mean squared radii
of $\Lambda (1405)$ are $\langle r^{2} \rangle _{\rm B} = 0.783 -
0.186 i \fm ^{2}$ and $\langle r^{2} \rangle _{\rm S} = - 1.097 +
0.662 i \fm ^{2}$, respectively~\cite{Sekihara:2008qk}, while $\sigma$
is a compact object with the mean squared scalar radius $\langle r^{2}
\rangle = (0.19 \pm 0.02) - (0.06 \pm 0.02) i \fm
^{2}$~\cite{Albaladejo:2012te}. However, due to the decay process, the
mean squared radius of the resonance state is evaluated as a complex
number, whose interpretation is not straightforward.  In
Ref.~\cite{Sekihara:2012xp} we have overcome this difficulty by using
the finite volume effect on the quasi-bound state.  Intuitively,
hadrons with a large (small) spatial size are sensitive (insensitive)
to the finite volume effect.  In the following, we quantitatively
formulate this picture using the mass shift in the finite volume.

\section{Structure of dynamically generated states with finite volume
  effect}

Here we investigate structures of dynamically generated hadronic
resonances obtained from the Bethe-Salpeter equation for the two-body
scattering amplitude $T_{ij}$ in an algebraic form:
\begin{equation}
  T_{ij} (s) = V_{ij} (s) + \sum _{k} V_{ik} (s) G_{k} (s) T_{kj} (s) 
  = \sum _{k} {(1 - V G )^{-1}}_{ik} V_{kj} ,
  \label{eq:BS}
\end{equation}
where $i$, $j$, and $k$ are channel indices, $s$ is the Mandelstam
variable, $V$ is the interaction kernel to be fixed later, and $G$ is
the two-body loop integral.  The resonances appear as poles in the
scattering amplitude as follows:
\begin{equation}
  T_{ij} (s) = \frac{g _{i} g_{j}}{s - s_{\rm pole}} 
  + T_{ij}^{\rm BG} (s) . 
  \label{eq:amp_pole}
\end{equation}
Here $g_{i}$ is coupling constant of the resonances to the two-body
state in channel $i$ and $T_{ij}^{\rm BG} (s)$ represents a background
contribution.  The coupling constant contains information on the
structure of the resonances, and it is discussed in
Ref.~\cite{Hyodo:2011qc} that the coupling constant $g_{i}$ is related
to the compositeness for the resonance with respect to the channel
$i$, $X_{i}$, which is the amount of two-body states composing the
resonance, in the following manner:
\begin{equation}
  X_{i} = - g_{i}^{2} \frac{d G_{i}}{d s} (s_{\rm pole}) .
  \label{eq:comp}
\end{equation}
The compositeness $X_{i}$ approaches unity if the state is dominated
by the two-body component in channel $i$, while it becomes zero if the
state does not contain the two-body component in channel $i$.

Next we consider the finite volume effect in a spatial box with the
periodic boundary condition of size $L$.  In the finite volume, the
momentum in the loop integral is discretized as $\bm{q} = 2 \pi \bm{n}
/ L$ ($\bm{n} \in \mathbb{Z}^{3}$)~\cite{Luscher:1985dn}.  Then, if a
stable bound state is put into a finite box, the mass of the bound
state $M_{\rm B}$ is changed due to the finite volume effect, and its
mass shift is predicted as~\cite{Luscher:1985dn}:
\begin{equation}
  \Delta M_{\rm B} ( L ) 
  = -\frac{3 g^{2}}{8 \pi M_{\rm B}^{2} L}
  \exp \left [ - \gamma L \right ] 
  + \mathcal{O}(e^{-\sqrt{2} \gamma L}), 
  \quad 
  \gamma \equiv \frac{\sqrt{- \lambda 
      (M_{\rm B}^{2}, \, m^{2}, \, M^{2})}}
  {2 M_{\rm B}} .
  \label{eq:mass-shift} 
\end{equation}
with the masses of two constituents $m$ and $M$.  Since this mass
shift formula contains the coupling constant squared $g^{2}$, which is
equivalent to the residue in Eq.~\eqref{eq:amp_pole}, the mass shift
formula brings a possibility to determine the structure of the bound
state with the finite volume effect.  Namely, putting a bound state
generated in Eq.~\eqref{eq:BS} into a finite box with the
momentum-discretized loop integral and measuring the mass shift
$\Delta M_{\rm B}(L)$, one can deduce the coupling constant from the
finite volume effect as:
\begin{equation}
  g_{\rm FV} = 
  \sqrt{\frac{\Delta M_{\rm B}(L)}
    {- 3 / (8 \pi M_{\rm B}^{2} L) 
      \exp \left [ - \gamma L \right ]}} , 
  \label{eq:gFV}
\end{equation}
and further deduce the compositeness from the finite volume effect as:
\begin{equation}
  X_{\rm FV} = - g_{\rm FV}^{2} 
  \frac{dG}{ds} (M_{\rm B}^{2}) . 
  \label{eq:XFV} 
\end{equation}
Then, since the mean squared distance between two constituents for the
bound state is evaluated as $X / (4 \mu B_{\rm E})$ with the reduced
mass $\mu$ for the small binding energy $B_{\rm E}$, we define the
mean squared distance from the finite volume effect as:
\begin{equation}
  \langle r^{2} \rangle _{\rm FV}
  = \frac{X_{\rm FV}}{4\mu B_{\rm{E}}} .
  \label{eq:RFV}
\end{equation}
We emphasize that our procedure can be easily applied to Feshbach
resonances with finite widths, which is a quasi-bound state of a
higher energy channel embedded in the continuum of a lower channel.
For Feshbach resonances, by putting the higher channel into a finite
box while the lower channel being unchanged and identifying the real
part of the resonance pole as the bound state mass $M_{\rm
  B}=\text{Re}[W_{\rm pole}]$, we can deduce the coupling constant,
compositeness, and mean squared distance with respect to the higher
channel.  Furthermore, it is important that we define the real-valued
distance between constituents for the resonance states with respect to
the closed channel.

\begin{figure}[!t]
  \centering
  \begin{tabular}{cc}
    \Psfig{0.49\textwidth}{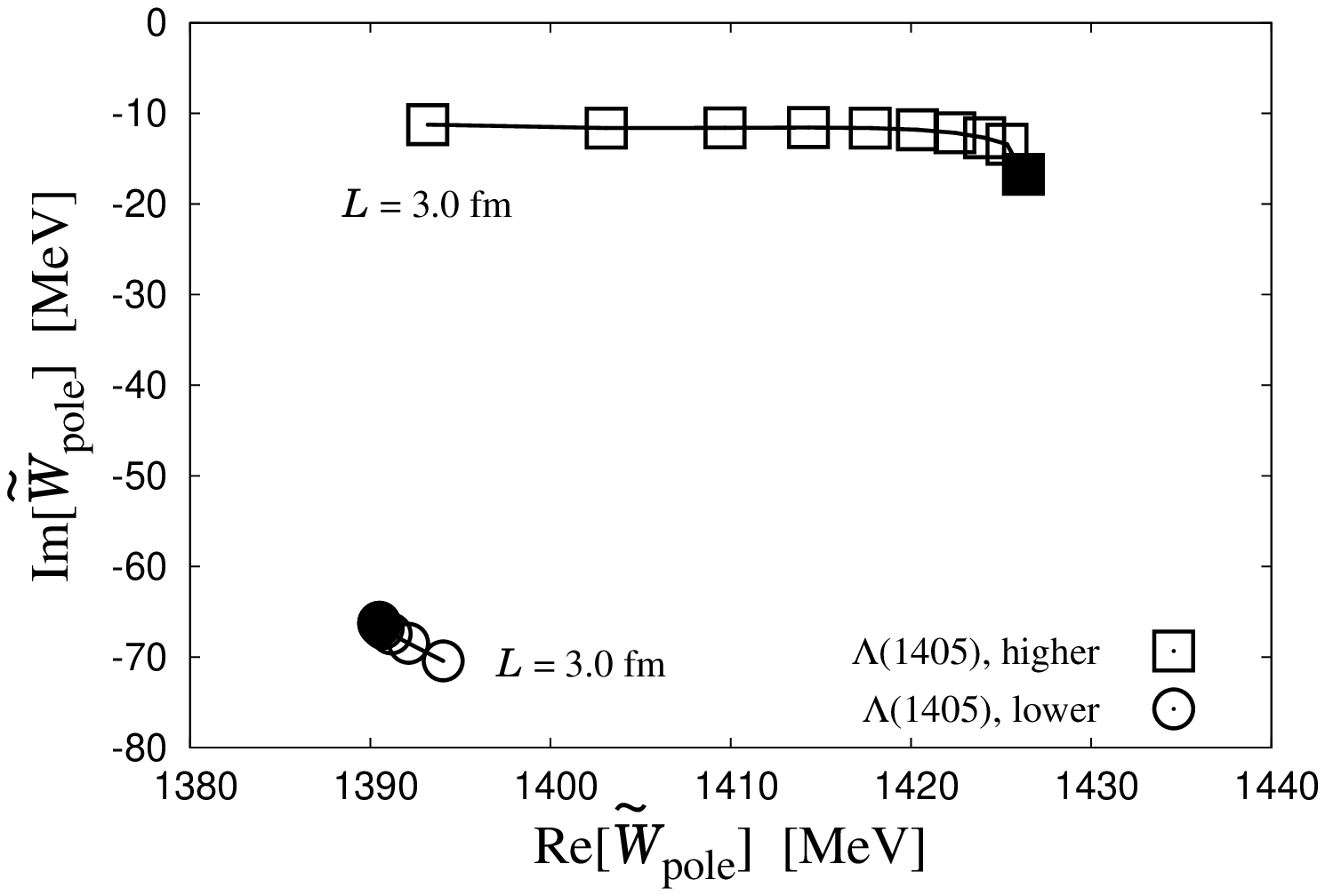} & 
    \Psfig{0.47\textwidth}{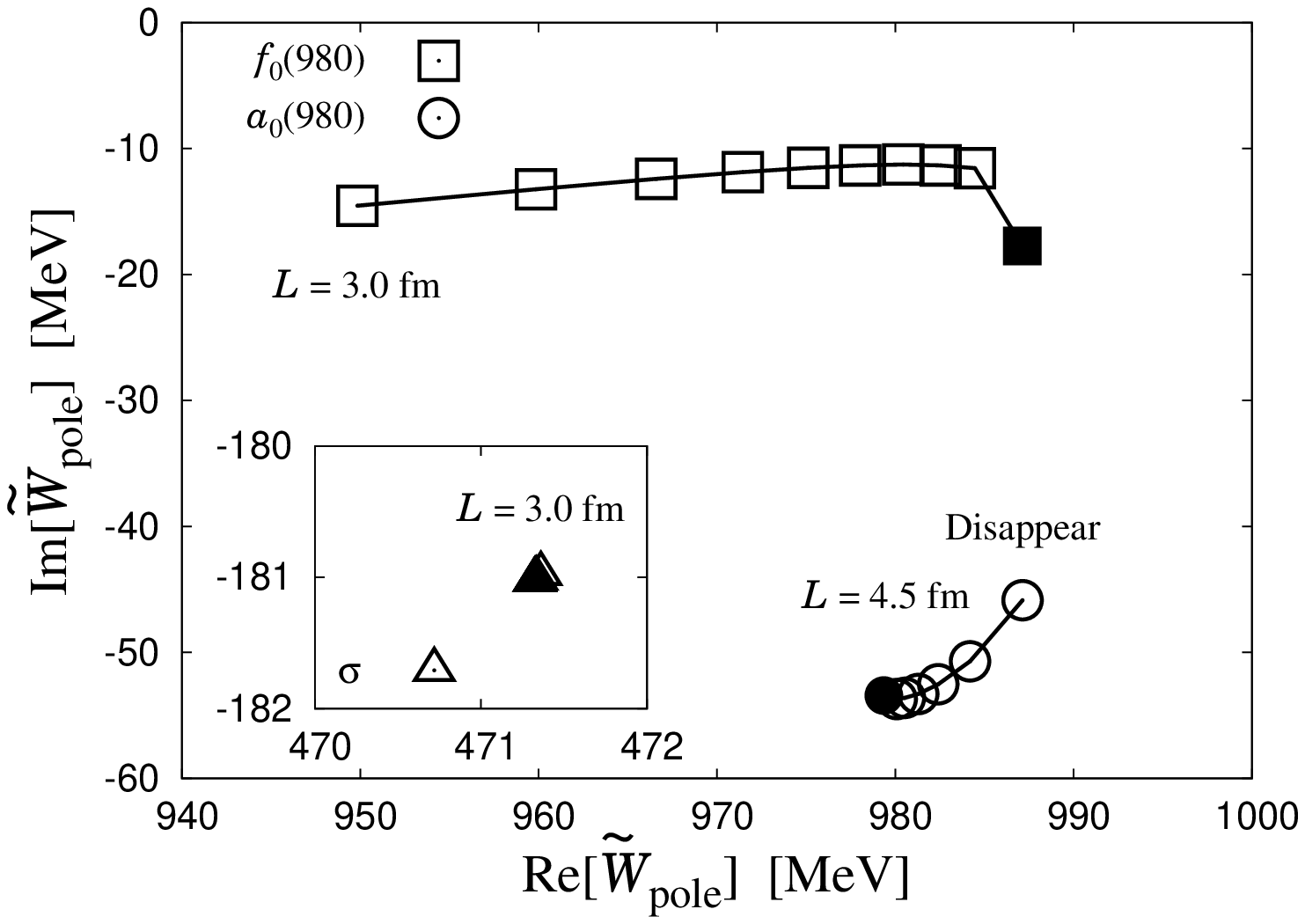} 
  \end{tabular}
  \caption{Behavior of the resonance pole positions for two $\Lambda
    (1405)$ (left) and three scalar mesons (right) in the complex
    energy plane~\cite{Sekihara:2012xp}.  Here filled symbols indicate
    the pole positions in infinite volume and open symbols are plotted
    in interval $0.5 \fm$ with respect to the box size $L$ from $L=3.0
    \fm$ to $L=7.0 \fm$.  }
  \label{fig:2}
\end{figure}

Let us now utilize our procedure for physical resonances in the chiral
unitary approach.  Here we discuss $\Lambda (1405)$ and the scalar
mesons $\sigma$, $f_{0}(980)$, and $a_{0}(980)$ in coupled-channels
problems with the the lowest-order $s$-wave chiral interaction.
Details of the model setup and the model parameters are given in
Ref.~\cite{Sekihara:2012xp}.  Our model in infinite volume dynamically
generates resonance poles of two $\Lambda (1405)$ and the scalar
mesons, which are shown as the filled symbols in Fig.~\ref{fig:2}.
Then, we put the $\bar{K}N$ and $K\bar{K}$ channels into finite boxes
with other channels being unchanged to extract the spatial size of
these channels.  Turning on the finite volume effect on the $\bar{K}N$
and $K\bar{K}$ channels, we observe shifts of the resonance pole
positions according to the box size $L$, which are plotted as open
symbols in interval $0.5 \fm$ from $L=3.0 \fm$ to $L=7.0 \fm$ in
Fig.~\ref{fig:2}.  From the figure, we can find that the poles of
higher $\Lambda (1405)$ and $f_{0}(980)$ move to lower energies while
other poles do not show downward mass shift.  This result implies that
higher $\Lambda (1405)$ and $f_{0}(980)$ have large $\bar{K}N$ and
$K\bar{K}$ components, respectively, but lower $\Lambda (1405)$,
$\sigma$, and $a_{0}(980)$ are not dominated by the $\bar{K}N$ nor
$K\bar{K}$ component.  From the mass shift of higher $\Lambda (1405)$
and $f_{0}(980)$ due to the finite volume effect, we can extract the
coupling constant via Eq.~\eqref{eq:gFV}, and further the
compositeness and the mean squared distance between two constituents
for the resonances as real values via Eqs.~\eqref{eq:XFV} and
\eqref{eq:RFV}, respectively.  The results are listed in
Table~\ref{tab:2}, and one can see that these resonances have
nearly-unity $\bar{K}N$ and $K\bar{K}$ compositeness with large
spatial extent between two constituents beyond the typical hadronic
size $\lesssim 0.8 \fm$.

\begin{table}
  \caption{Properties of $\Lambda (1405)$ and $f_{0}(980)$ with finite
    volume effect~\cite{Sekihara:2012xp}.  
  }
  \label{tab:2}
  \centering
  \begin{tabular}{ccccc}
    \hline \hline
    \multicolumn{2}{c}{$\Lambda (1405)$, higher pole} & 
    & \multicolumn{2}{c}{$f_{0}(980)$} 
    \\
    \hline
    $g_{\bar{K} N, \rm FV}$ & $4.6$ -- $5.2 \gev$ & 
    & $g_{K \bar{K}, \rm FV}$ & $2.7$  -- $3.1 \gev$ 
    \\
    $X_{\bar{K}N,\rm{FV}}$ & $0.82$ -- $1.03$ & 
    & $X_{K\bar{K},\rm FV}$ & $0.73$ -- $0.97$
    \\
    $\sqrt{\langle r^{2} \rangle _{\bar{K} N, \rm FV}}$ & $1.7$ -- $1.9 \fm $ & 
    & $\sqrt{\langle r^{2} \rangle _{K \bar{K}, \rm FV}}$ & $2.6$ -- $3.0 \fm $ 
    \\
    $\sqrt{\langle R^{2} \rangle _{\rm size, FV}}$ & $1.1$ -- $1.2 \fm $ & 
    & $\sqrt{\langle R^{2} \rangle _{\rm size, FV}}$ & $1.4$ -- $1.6 \fm $
    \\
    \hline \hline
  \end{tabular}
\end{table}

Furthermore, with spatial structures of constituents taken into
account, we can estimate the root mean squared radii of the resonances
$\sqrt{\langle R^{2} \rangle _{\rm size, FV}}$ from a kinematical
consideration, which results in $1.1$--$1.2 \fm$ for higher $\Lambda
(1405)$ and $1.4$--$1.6 \fm$ for $f_{0}(980)$.  Therefore, the root
mean squared radii of higher $\Lambda (1405)$ and $f_{0}(980)$ are
larger than the typical hadronic scale $\lesssim 0.8 \fm$.  

Finally we mention that the resonance mass may have uncertainties of
the half-width, which is a subtle problem because the mean distance of
the state is sensitive to the binding energy as in Eq.~\eqref{eq:RFV}.
This point has been also discussed in Ref.~\cite{Sekihara:2012xp} by
changing the mass $M_{\rm B} = \text{Re}[ W_{\rm pole} ] \to
\text{Re}[ W_{\rm pole} ] - \Gamma /2$, and it is found that, although
the mean squared distance decreases with the scaling $\propto 1/B_{\rm
  E}$, the compositeness stays similar values regardless of the mass
uncertainties.

\section{Summary}

We have investigated the structures of the dynamically generated
hadronic resonances in the chiral unitary approach from the finite
volume effect.  For this purpose we have established the relation
between the mass shift of a Feshbach resonance coming from the finite
volume effect and the mean squared distance between two constituents
of the resonance.  Especially we can define the real-valued size of
the resonance in a given closed channel from response to the finite
volume effect on the channel.
Utilizing this relation to the physical hadrons, we investigate the
structure of $\Lambda (1405)$ and scalar mesons $\sigma$,
$f_{0}(980)$, and $a_{0}(980)$ with respect to the $\bar{K}N$ and
$K\bar{K}$ components.  We have found that the poles of the higher
$\Lambda (1405)$ and $f_{0}(980)$ move to lower energies while other
poles do not show downward mass shift, which implies large $\bar{K}N$
and $K\bar{K}$ components inside higher $\Lambda (1405)$ and
$f_{0}(980)$, respectively.  The compositeness of $\bar{K}N$
($K\bar{K}$) inside $\Lambda (1405)$ [$f_{0}(980)$] is evaluated as
$0.82$ -- $1.03$ ($0.73$ -- $0.97$) and the mean distance between two
constituents is $1.7$--$1.9 \fm$ ($2.6$--$3.0 \fm$).  Furthermore,
from a kinematical consideration the root mean squared radii of the
resonances are evaluated as $1.1$--$1.2 \fm$ for higher $\Lambda
(1405)$ and $1.4$--$1.6 \fm$ for $f_{0}(980)$.  As a consequence, both
the root mean squared distances and radii for $\Lambda (1405)$ and
$f_{0}(980)$ are larger than the typical hadronic scale $\lesssim 0.8
\fm$.

\end{document}